\documentclass[12pt,oneside]{article}
\usepackage{amsfonts,amssymb,graphicx,psfig}
\usepackage{amsmath}
\usepackage{subfigure}
\def\no{\noindent}
\def\be{\begin{equation}}
\def\ee{\end{equation}}
\def\bea{\begin{eqnarray}}
\def\eea{\end{eqnarray}}

\def\<{\langle}
\def\>{\rangle}
\def\~{\tilde}

\def\a{\alpha}

\def\g{\gamma}

\def\cosh{{\rm cosh}}

\def\Var{{\rm Var}}

\def\ds{\displaystyle}

\newcommand{\R}{\mathbb R}

\newcommand{\Z}{\mathbb Z}

\newcommand{\ulimit}[1]{\underset{#1}{\longrightarrow}}

\newtheorem{proposition}{Proposition}
\newtheorem{theorem}{Theorem}

\newtheorem{lemma}{Lemma}

\newenvironment{proof}{\no{\it Proof:}}{\hfill$\square$\vskip.5cm}
\newenvironment{proofof}[1]{\no{\it Proof of #1:}}{\hfill$\square$\vskip.5cm}

\begin{document}

\begin{center}
\vspace{1truecm}
{\bf\sc\Large bipartite mean field spin systems.\\ 
existence and solution }\\
\vspace{1cm}
{Ignacio Gallo, \quad Pierluigi Contucci}\\
\vspace{.5cm}
{\small Dipartimento di Matematica, Universit\`a di Bologna,\\ {e-mail: {\em gallo@dm.unibo.it, contucci@dm.unibo.it}}} \\
\end{center}
\vskip 1truecm
\begin{abstract}\noindent
A mean field spin system consisting two interacting groups each with homogeneous
interaction coefficients is introduced and studied. Existence of the thermodynamic limit
is shown by an asymptotic sub-addittivity method and factorization of correlation functions
is proved almost everywhere. The free energy solution of the model is obtained by upper and lower
bounds and by showing that their difference vanishes for large volumes.
\end{abstract}

\section{Introduction}

In this work we consider the problem of characterizing the equilibrium statistical mechanics
of an interacting system of two set of spins. We aim to tackle the most general case of a two
population system in the mean field approximation.


Mean field two-population models have been useful since
the study of metamagnets which started in the '70s (see \cite{cohen, cohen.kincaid}), and have been
encountered very recently in the study
of loss of gibbsianness for a system whose evolution is described by Glauber dynamics (see \cite{kulske}).
In Refs \cite{cohen, cohen.kincaid} a two-population
mean-field model is used as an approximation to a bipartite
lattice assumed to describe an antiferromagnetic system, and is found to reproduce qualitatively the
expected phase
transitions, which are then studied at criticality. In Ref. \cite{kulske}, instead,
particles are subject to a
time-evolving random field which acts on particles by partioning them in two groups, leading to a mean field
model mathematically
analogous to the former. In this case a characterization of the whole phase diagram is provided.

The systems considered by both works can be seen as a restriction of a more general model,
which is presented here, and which
arises naturally as an interacting mixture of two systems of Curie-Weiss type.

Our results can be summarised as follows. After introducing the model we show in section
3 that it is well posed by showing that its thermodynamic limit exists. The result is non trivial
because sub-additivity is not met at finite volume. In section 4 we show that the system 
fulfills a factorization property for the correlation functions which reduces the equilibrium state 
to only two degrees of freedom the equilibrium state. The method is conceptually similar to the one developed by
Guerra in \cite{aod} to derive identities for the overlap distributions in the Sherrington and Kirkpatrick
model.

We also derive the pressure of the model by rigorous methods developed in the recent
study of mean field spin glasses (see \cite{guerrarev} for a review). It is interesting to
notice that though very simple,
our model encompasses a range of regimes that do not admit solution by the
elegant interpolation method used in the celebrated existence result of the Sherrington and Kirkpatrick model \cite{guerra}. 
This is due to the lack of positivity of the quadratic form describing
the considered interaction.
Nevertheless we are able to solve the model exactly, section 5, using the lower bound provided by the Gibbs
variational principle, and thanks to a further bound given by
a partitioning of the configuration space, itself originally devised in the study of spin glasses
(see \cite{guerrarev, desanctis, guerrarev2}).

As in the classical Curie-Weiss model, the exact solution is provided in an implicit form;
for our system, however, we find two equations of state, which are coupled as well as trascendental,
and this makes the full characterization of all the possible regimes highly non-trivial:
a robust numerical analysis becomes essential and can be found in \cite{cgm}, where an application to social sciences
is considered.

Some aspects of the regimes can nonetheless be studied analytically, and this is done in section
\ref{maxima}, while a global study of the phase diagram for our model is left to be carried on
in a future work.

\section{The Model}\label{model}

Our model is defined by the Hamiltonian
\begin{equation}\label{Ham}
    H(\sigma)=-\frac{1}{2N}\sum_{i, j=1}^{N}J_{ij}\sigma_{i}\sigma_{j}-\sum_{i}h_{i}\sigma_{i} \;.
\end{equation}

We consider Ising spins, $\sigma_i=\pm 1$, and symmetric interactions $J_{i,j}$.
We divide the particles $P=\{1,2,3,...,N\}$ into 2 types $A$ and $B$ with
$A \cup B = P$, $A \cap B = \emptyset$, and sizes $N_1=|A|$ and $N_2=|B|$,
where $N_1+N_2=N$. Given two particles $i$ and $j$, their mutual interaction
parameter $J_{ij}$ depends on the subset they belong to, as specified by the matrix
\begin{displaymath}
         \begin{array}{ll}
                \\
                N_1 \left\{ \begin{array}{ll}
                \\
                                  \end{array}  \right.
                \\
                N_2 \left\{ \begin{array}{ll}
                \\
        	\\
                \\
                                  \end{array}  \right.
         \end{array}
          \!\!\!\!\!\!\!\!
         \begin{array}{ll}
                \quad
                 \overbrace{\qquad }^{\textrm{$N_1$}}
                 \overbrace{\qquad \qquad}^{\textrm{$N_2$}}
                  \\
                 \left(\begin{array}{c|ccc}
                               \mathbf{ J_{11}}  &  & \mathbf{ J_{12}}
                                \\
                                 \hline
                                &  &  &
                                \\
                                \mathbf{ J^{*}_{12}} &  & \mathbf{ J_{22}}
                                \\
                                &   &   &
                                \\
                      \end{array}\right)
               \end{array}
\end{displaymath}

\no where each matrix block has constant elements:
$J_{11}$ and $J_{22}$ tune the interactions within each of
the two types, and $J_{12}$ controls the interaction between two
particles of different types. In view of the applications
considered in the introduction, we assume $J_{11}>0$ and
$J_{22}>0$, whereas $J_{12}$ can be either positive or negative.

Analogously, the field $h_i$ takes two values $h_1$ and $h_2$, depending on the type of $i$, as described by the  following vector:
\begin{displaymath}
         \begin{array}{ll}
                N_1 \left\{ \begin{array}{ll}
                                      \\
                                   \end{array}  \right.
                                        \\
                N_2 \left\{ \begin{array}{ll}
                                        \\
                                          \\
                                         \\
			     \end{array}  \right.
	       \!\!\!\!\!\!
	\end{array}
	\!\!\!\!\!\!
	\left(\begin{array}{ccc|c}
     			\mathbf{h_{1}}
 			\\
			\hline
			\\
			\mathbf{h_{2}}
			\\
			\\
		\end{array}\right)
\end{displaymath}


By introducing the magnetization of a subset $S$ as
\begin{displaymath}
m_S(\sigma) \; = \; \frac{1}{|S|}\sum_{i\in S}\sigma_i
\end{displaymath}

\noindent and indicating by $m_1$ and $m_2$ the magnetizations within
the subsets $A$ and $B$ and by
$\displaystyle{\alpha=\frac{N_1}{N}}$ the relative size of subset A
on the whole, we may easily express the Hamiltonian per particle as
\begin{equation}\label{Ham}
\frac{H(\sigma)}{N}=-\frac{1}{2}\left[J_{11}\alpha^2m_1^2\!+\! 2J_{12}\alpha(1\!-\!\alpha)m_1m_2\!+\!
J_{22}(1\!-\!\alpha)^2m_2^2\right]   - h_1\alpha m_1-h_2(1-\alpha)m_2
\end{equation}

The usual statistical mechanics framework defines the equilibrium value of an observable $f(\sigma)$ as the
average with respect to the {\it Gibbs distribution} defined by the Hamiltonian. We call this average the {\it Gibbs state}
for $f(\sigma)$, and write it explicitly as:
\begin{displaymath}
\< \, f \, \> \; = \; \frac{\sum_{ \sigma} f(\sigma) \; e^{\,-H(\sigma)}    }{\sum_\sigma e^{\,-H(\sigma)}} \; .
\end{displaymath}

The main observable for our model is the average of a spin configuration,
i.e. the {\it magnetization}, $m(\sigma)$, which explicitly reads:
\begin{displaymath}
m(\sigma)=\frac{1}{N}\sum_{i=1}^N \sigma_i.
\end{displaymath}

Our quantity of interest is therefore $\<m\>$: to find it, as well as the moments of many other observables,
statistical mechanics leads us to consider the pressure function:
\begin{displaymath}
p_N \; = \; \frac{1}{N}\log \sum_{\sigma}e^{-H(\sigma)} \; .
\end{displaymath}

It is easy to verify that, once it's been derived exactly, the pressure is capable
of generating the Gibbs state for the magnetization as
$$
\< m \> = \alpha \frac{\partial p_N}{\partial h_1} + (1-\alpha) \frac{\partial p_N}{\partial h_2}.
$$

In order to simplify the analytical study of the model it is useful to observe that
the Hamiltonian is left invariant by the action of a group of transformations,
so that only a subspace of parameter space needs to be considered.

The symmetry group is described by $G=\Z_2 \times \Z_2 \times \Z_2$.

We can represent a point in our parameter space as $(\mathbf{m},\mathbf{J},\mathbf{h}, \mathbf{\hat\alpha})$, where
\[
\mathbf{m}=\left(\begin{array}{c}
         m_1
         \\
         m_2
    \end{array}\right),
    \quad
\mathbf{J}=\left(\begin{array}{cc}
         J_{11}&J_{12}
         \\
         J_{12}&J_{22}
    \end{array}\right),
    \quad
\mathbf{h}=\left(\begin{array}{c}
         h_1
         \\
         h_2
    \end{array}\right),
    \quad
\mathbf{\hat\alpha}=\left(\begin{array}{cc}
         \alpha&0
         \\
         0&1-\alpha
    \end{array}\right).
\]

Therefore, given the limitations on the values of our parameters, the whole parameter space is given by
$S=[-1,1]^2\times\R^2 \times\R_+ \times \R^2 \times [0,1]$.

If we consider the representation of $G$ given by the
$8$ matrices
\[
\left(\begin{array}{cc}
         \epsilon_1&0
         \\
         0&\epsilon_2
    \end{array}\right),
    \quad
         \epsilon_i=+1 \textrm{ or } -1
\quad
\textrm{ and }
\quad
\left(\begin{array}{cc}
         0 & \eta_1
         \\
         \eta_2 & 0
    \end{array}\right),
    \quad
         \eta_i=+1 \textrm{ or } -1
\]

\noindent we can immediately realize that $G$ is a symmetry from the
Hamiltonian representation:
\[
\frac{H(\mathbf{m},\mathbf{J},\mathbf{h}, \mathbf{\hat\alpha)}}{N} \;=
\; - \frac{1}{2}\<\mathbf{\hat\alpha} \, \mathbf{m}, \mathbf{J} \mathbf{\hat\alpha} \, \mathbf{m}\>
- \< \mathbf{h},  \mathbf{\hat\alpha} \, \mathbf{m} \>.
\]

%

%


%


\section{Existence of the thermodynamic limit}

We shall prove that our model admits a thermodynamic limit by exploiting an existence theorem
provided for mean field models in \cite{bcg}: the result states that the existence of the pressure
per particle for large volumes is guaranteed by a monotonicity condition on the equilibrium
state of the Hamiltonian. Such a result proves to be quite useful when the condition of
convexity introduced by the interpolation method \cite{guerra,guerrarev} doesn't apply due to
lack of positivity of the quadratic form representing the interactions.
%
We therefore prove the existence of the thermodynamic limit independently of an exact solution. Such
a line of enquiry is pursued in view of further refinements of our model, that shall possibly involve
random interactions of spin glass or random graph type, and that might or might not come with
an exact expression for the pressure.
\begin{proposition}\label{existence}
There exists a function $p$ of the parameters $(\alpha, J_{1,1}, J_{1,2}, J_{2,2}, h_1, h_2)$ such that
$$
\lim_{N \rightarrow \infty} p_N= p \; .
$$
\end{proposition}
\vskip .1 in
The previous proposition is proved with a series of lemmas.
Theorem 1 in \cite{bcg} states that given an Hamiltonian $H_N$ and its associated equilibrium state
$\omega_N$ the model admits a thermodynamic limit whenever the physical condition
\begin{equation}\label{additivity}
\omega_N (H_N) \geqslant \omega_N (H_{K_1}) + \omega_N (H_{K_2}), \qquad K_1 + K_2= N,
\end{equation}
\no is verified.

We proceed by first verifying this condition for a working Hamiltonian $\tilde H_N$, and then showing
that its pressure $\tilde p_N$ tends to our original pressure $p_N$ as $N$ increases. We choose $\tilde H_N$
in such a way that the condition (\ref{additivity}) is verified as an equality.

Our working Hamiltonian $\tilde H_N$ is defined as follows:
$$
\tilde H_N=\tilde H_N^{(1)}+\tilde H_N^{(12)}+\tilde H_N^{(2)},
$$

\no where
$$
\tilde H_N^{(1)}=\alpha J_{11} \frac{1}{\alpha N - 1 }\sum_{i \neq j=1 \dots N_1} \xi_i \xi_j, \qquad
\tilde H_N^{(2)}=(1-\alpha) J_{22}\frac{1}{(1-\alpha)N-1}\sum_{i \neq j=1 \dots N_2} \eta_i \eta_j,
$$
$$
\tilde H_N^{(12)}=\frac{1}{N}J_{12}\sum_{i=1 \dots N_1 \atop j=1 \dots N_2} \xi_i \eta_j.
$$


\begin{lemma}\label{workinglimit}
There exists a function $\tilde p$ such that
$$
\lim_{N \rightarrow \infty} \tilde p_N= \tilde p
$$
\end{lemma}

\vskip .1 in
\begin{proof}
By definition of $H_N^{(1)}$ and by the invariance of $\omega_N$ with respect to spin permutations,
$$
\omega_N(\tilde H_N^{(1)}) = \omega(\alpha J_1 \frac{1}{\alpha N - 1 }\sum_{i\neq j =1^N_1} \xi_i \xi_j )=
\alpha J_1 \frac{(\alpha N-1) \alpha N}{\alpha N -1}\omega_N(\xi_1 \xi_2) = N \alpha ^2 J_1 \omega (\xi_1 \xi_2).
$$

We can find a similar form for $\tilde H^{(2)}$ and $\tilde H^{(12)}$, which implies that for any two positive integers
$K_1+K_2=N$ we have
$$
\omega(\tilde H_N)=\omega( \tilde H_{K_1}+\tilde H_{K_2}),
$$
\no which verifies (\ref{additivity}) and proves our Lemma.
\end{proof}

The following two Lemmas show that the difference between $H_N$ and $\tilde H_N$ is
thermodynamically negligible and as a consequence their pressures coincide in the thermodynamic
limit.

For convenience we shall re-express our Hamiltonian in the following way:
$$
H_N=H_N^{(1)}+H_N^{(12)}+H_N^{(2)}
$$

\no where we define
$$
H_N^{(1)}=\frac{1}{N}J_{11}\sum_{i, j=1 \dots N_1} \xi_i \xi_j, \qquad
H_N^{(2)}=\frac{1}{N}J_{22}\sum_{i, j=1 \dots N_2} \eta_i \eta_j,
$$
$$
H_N^{(12)}=\frac{1}{N}J_{12} \ \sum_{i=1 \dots N_1 \atop j=1 \dots N_2} \xi_i \eta_j \; .
$$

\begin{lemma}\label{order}

\begin{equation}
H_N = \tilde H_N + O(1)
\end{equation}

\no i.e.
$$
\lim_{N \rightarrow \infty}  \frac{H_N}{N} = \lim_{N \rightarrow \infty} \frac{\tilde H_N}{N}
$$

\end{lemma}

\begin{proof}
$$
H_N^{(1)}=\frac{1}{N}J_{11}\sum_{i, j=1 \dots N_1} \xi_i \xi_j
=\frac{N_1-1}{N} J_{11} \frac{1}{N_1 - 1 }\sum_{i \neq j=1 \dots N_1} \xi_i \xi_j + \frac{1}{N}J_{11}\sum_{i=1 \dots N_1} \xi_i \xi_i,
$$

\no and since $\ds {\alpha=\frac{N_1}{N}}$
\begin{eqnarray*}
& = &\frac{\alpha N -1}{\alpha N} \alpha J_{11} \frac{1}{\alpha N - 1 }\sum_{i \neq j=1 \dots N_1} \xi_i \xi_j + \alpha J_{11}=
\\
& = &  \alpha J_{11} \frac{1}{\alpha N - 1 }\sum_{i \neq j=1 \dots N_1} \xi_i \xi_j
-  \alpha J_{11} \frac{1}{\alpha N(\alpha N - 1)}\sum_{i \neq j=1 \dots N_1} \xi_i \xi_j
+ \alpha J_{11},
\end{eqnarray*}

\no and so
$$
H^{(1)}_N = \tilde H^{(1)}_N + O(1).
$$

We can similarly get estimates for $H^{(1)}_N$ and $H^{(12)}_N$ in terms of $\tilde H^{(1)}_N$ and $\tilde H^{(12)}_N$, which implies
$$
H_N = \tilde H_N + O(1).
$$
\end{proof}

\begin{lemma}\label{presham}
Say $\ds{p_N=\frac{1}{N} \ln Z_N }$, and say $h_N(\sigma)=\ds{\frac{H_N(\sigma)}{N}   }$.
Define $\tilde Z$, $\tilde p_N$ and $\tilde h_N$ in an analogous way.

Define
\begin{equation}\label{norm}
	k_N = \ds{\|h_N-\tilde h_N \|=\sup_{\sigma \in \{-1,+1\}^{N}} \{ |h_N(\sigma)- \tilde h_N(\sigma) | \} < \infty  }.
\end{equation}

Then
$$
|p_N- \tilde p_N| \leqslant \|h_N- \tilde h_N\| \; .
$$
\end{lemma}

\begin{proof}
\begin{eqnarray*}
p_N-\tilde p_N&=& \frac{1}{N} \ln Z_N - \frac{1}{N} \ln \tilde Z_N = \frac{1}{N} \ln \frac{Z_N}{\tilde Z_N}
\\
&=& \frac{1}{N} \ln \frac{\sum_{\sigma} e^{-H_N(\sigma)} }{\sum_{\sigma}e^{-\tilde H_N(\sigma)}}
\leqslant \frac{1}{N} \ln \frac{\sum_{\sigma} e^{-H_N(\sigma)}}{\sum_{\sigma}e^{-N(h_N(\sigma) + k_N)}}=
\\
&=& \frac{1}{N} \ln \frac{\sum_{\sigma} e^{-H_N(\sigma)}}{e^{-Nk_N}\sum_{\sigma}e^{-N h_N(\sigma)}}= \frac{1}{N} \ln e^{Nk_N}
= k_N = \|h_N-\tilde h_N\|
\end{eqnarray*}

\no where the inequality follows from the definition of $k_N$ in (\ref{norm}) and from monotonicity of the exponential and logarithmic functions.
The inequality for $\tilde p_N-p_N$ is obtained in a similar fashion.
\end{proof}

We are now ready to prove the main result for this section:

\begin{proofof}{Proposition \ref{existence}}
The existence of the thermodynamic limit follows from our Lemmas. Indeed, since
by Lemma \ref{workinglimit} the limit for $\tilde p_N$ exists, Lemma \ref{presham}
and Lemma \ref{order} tell us that
$$
\lim_{N \rightarrow \infty} |p_N-\tilde p_N | \leqslant \lim_{N \rightarrow \infty} \|h_N-\tilde h _N \| = 0,
$$
\no implying our result.
\end{proofof}

\section{Factorization properties}

In this section we shall prove that the correlation functions of our model factorize completely
in the thermodynamic limit, for almost every choice of parameters. This implies that
all the thermodynamic properties of the system can be described by the magnetizations
$m_1$ and $m_2$ of the two subsets $A$ and $B$ defined in Section \ref{model}. Indeed, the exact solution
of the model,
to be derived in the next section, comes as two coupled equations of state for $m_1$
and $m_2$.

\begin{proposition}
$$
\lim_{N \rightarrow \infty} \big ( \omega_N(\sigma_i\sigma_j)-\omega_N(\sigma_i)\omega_N(\sigma_j) \big ) =0
$$

\no for almost every choice of parameters  $(\alpha, J_{11},J_{12},J_{22},h_1, h_2)$, where $\sigma_i$, $\sigma_j$
are spins of any two particles in the system.
\end{proposition}

\begin{proof}
We recall the definition of the Hamiltonian per particle
\begin{equation*}
\frac{H_N(\sigma)}{N}=-\frac{1}{2}\left[J_{11}\alpha^2m_1^2\!+\! 2J_{12}\alpha(1\!-\!\alpha)m_1m_2\!+\!
J_{22}(1\!-\!\alpha)^2m_2^2\right]   - h_1\alpha m_1-h_2(1-\alpha)m_2,
\end{equation*}
and of the pressure per particle
$$
p_N=\frac{1}{N} \ln \sum_{\sigma} e^{-H_N(\sigma)}.
$$

By taking first and second partial derivatives of $p_N$ with respect to $h_1$ we get
$$
\frac{\partial p_N}{\partial h_1}=\frac{1}{N} \sum_{\sigma} \alpha N m_1(\sigma) \frac{e^{-H(\sigma)}}{Z_N}=\alpha \omega_N(m_1),
\qquad
\frac{\partial^2 p_N}{\partial \; h_1^2}=N \alpha^2 ( \omega_N(m_1^2)-\omega_N(m_1)^2).
$$

By using these relations we can bound above the integral with respect to $h_1$ of the fluctuations of $m_1$ in the Gibbs state:
\begin{eqnarray}\label{average var}
\Bigg |\int_{h_1^{(1)}}^{h_1^{(2)}} ( \omega_N(m_1^2)-\omega_N(m_1)^2 ) \; dh_1 \Bigg | & = &
\frac{1}{N \alpha^2} \Bigg | \int_{h_1^{(1)}}^{h_1^{(2)}}   \frac{\partial^2 p_N}{\partial h_1^2}   \; dh_1 \Bigg | =
\frac{1}{N \alpha^2} \Bigg | \int_{h_1^{(1)}}^{h_1^{(2)}} \frac{\partial p_N}{\partial \; h_1} \bigg |_{h_1^{(1)}}^{h_1^{(2)}} \Bigg | \leqslant
\nonumber  \\
& \leqslant & \frac{1}{N \alpha} \big ( \big | \omega_N(m_1)|_{h_1^{(2)}} \big | + \big | \omega_N(m_1)|_{h_1^{(1)}} \big | \big ) = O \big (\frac{1}{N} \big ).
\nonumber  \\
\end{eqnarray}

On the other hand, given any $\a \in (0,1)$ we have that
$$
\omega_N(m_1)=\frac{1}{\alpha}\frac{\partial p_N}{\partial h_1},
$$
and
$$
\omega_N(m_1^2)=\frac{2}{\alpha^2}\frac{\partial p_N}{\partial J_{11}},
$$
so, by convexity of the thermodynamic pressure $\ds p=\lim_{N \rightarrow \infty} p_N$,
both quantities $\ds \frac{\partial p_N}{\partial h_1}$ and $\ds \frac{\partial p_N}{\partial J_{11}}$ have well
defined thermodynamic limits almost everywhere.
This together with (\ref{average var}) implies that
\begin{equation}\label{ae m1}
\lim_{N \rightarrow \infty} ( \omega_N(m_1^2)-\omega_N(m_1)^2 ) = 0 \quad \textrm{a.e. in $h_1$, $J_{11}$}.
\end{equation}

In order to prove our statement we first consider spins of particles of type $A$, which we shall call $\xi_i$. Translation
invariance of the Gibbs measure tells us that
\begin{eqnarray}\label{xi i xi j}
\omega_N(m_1)&=&\omega_N(\frac{1}{N}\sum_{i=1}^{N_1} \xi_i )=\alpha \omega_N(\xi_1),
\nonumber \\
\omega_N(m_1^2)&=&\omega_N(\frac{1}{N^2}\sum_{i,j=1}^{N_1} \xi_i \xi_j )=
\omega_N(\frac{1}{N^2}\sum_{i\neq j=1}^{N_1} \xi_i \xi_j )+\omega_N(\frac{1}{N^2}\sum_{i = j=1}^{N_1} \xi_i \xi_j ) =
\alpha \frac{N_1-1}{N}\omega_N(\xi_1 \xi_2) + \frac{\alpha}{N}. \nonumber
\\
\end{eqnarray}

We have that (\ref{xi i xi j}) and (\ref{ae m1}) imply
\begin{equation}\label{part one}
\lim_{N \rightarrow \infty} \omega_N(\xi_i\xi_j)-\omega_N(\xi_i)\omega_N(\xi_j)=0,
\end{equation}

\no which verifies our statement for all couples of spins $i \neq j$ of type $A$ as defined
in section \ref{model} (the case $i=j$ verifies (\ref{part one}) trivially).

Working in strict analogy as above we also get
\begin{equation}\label{ae m2}
\lim_{N \rightarrow \infty} ( \omega_N(m_2^2)-\omega_N(m_2)^2 ) = 0 \quad \textrm{a.e. in $h_2$, $J_{22}$}.
\end{equation}
Furthermore, by defining $\Var_N(m_1)=\big (\omega_N(m_1^2)-\omega_N(m_1)^2 \big )$, and
analogously for $m_2$, we exploit (\ref{ae m1}) and (\ref{ae m2}), and
use the Cauchy-Schwartz inequality to get
\begin{equation}\label{ae m1 m2}
|\omega_N(m_1 m_2)-\omega_N(m_1)\omega_N(m_2)| \leqslant \sqrt{\Var(m_1)\Var(m_2)} \ {\ds{ \ulimit{\footnotesize N \rightarrow \infty}}} \ 0
\quad \textrm{a.e. in $J_{11}$, $J_{12}$, $J_{22}$, $h_1$, $h_2$}
\end{equation}

By using (\ref{ae m2}) and (\ref{ae m1 m2}) we can therefore verify 
statements which are analogous to \ref{part one}, but which concern
$\omega_N(\xi_i \eta_j)$ and $\omega_N(\eta_i \eta_j)$ where $\xi$ are spins of
type $A$ and $\eta$ are spins of type $B$.

We have thus proved our claim for any couple of spins in the global system.


\end{proof}

\section{Solution of the model}


We shall derive upper and lower bounds for the thermodynamic limit of the pressure. The lower bound is obtained through the standard entropic variational principle, while the upper bound
is derived by a decoupling strategy.
%

\subsection{Upper bound}

In order to find an upper bound for the pressure we shall divide the configuration space into a partition of
microstates of equal magnetization, following \cite{desanctis, guerrarev, guerrarev2}. Since
subset A consists of $N_1$ spins, its magnetization can take
exactly $N_1+1$ values, which are the elements of the set
$$
R_{N_1}=\Big \{-1, -1+\frac {1} {2 N_1}, \dots ,1 -\frac {1} {2 N_1}, 1 \Big \}.
$$

Clearly for every $m_1(\sigma)$ we have that
$$
\sum_{\mu_1 \in R_{N_1}} \delta_{m_1, \mu_1}=1,
$$

\no where $\delta_{x,y}$ is a Kronecker delta.
We can similarly define a set $R_{N_2}$, so we have that
\begin{eqnarray} \label{deltas}	
Z_N &= &\sum_\sigma   \exp \big\{ \frac{N}{2} (J_{11}\alpha^2m_1^2+ 2J_{12}\alpha(1-\alpha)
m_1m_2+J_{22}(1-\alpha)^2m_2^2) +  h_1N_1m_1 + h_2N_2m_2 \big \} ={}
\nonumber \\
     & = & \sum_\sigma \sum_{\mu_1 \in R_{N_1}  \atop \mu_2 \in R_{N_2} }
          \delta_{m_1, \mu_1} \delta_{m_2, \mu_2}
   \exp\big\{\frac{N}{2}(J_{11}\alpha^2m_1^2+ 2J_{12}\alpha(1-\alpha)m_1m_2+J_{22}(1-\alpha)^2m_2^2)+ {}
            \nonumber\\
 &  &  {} +  h_1N_1m_1 + h_2N_2m_2 \big \}.
\end{eqnarray}

Thanks to the Kronecker delta symbols,
we can substitute $m_1$ (the average of the spins within a configuration)
with the parameter $\mu_1$ (which is not coupled to the spin configurations)
in any convenient fashion, and the same holds for $m_2$ and $\mu_2$.



Therefore we can use the following relations in order to linearize all quadratic terms
appearing in the Hamiltonian
\begin{eqnarray*}
(m_1 - \mu_1)^2 & = & 0,
\nonumber\\
(m_2 - \mu_2)^2 & = & 0,
\nonumber\\
(m_1 - \mu_1)(m_2 - \mu_2) & = & 0.
\end{eqnarray*}


Once we've carried out these substitutions into (\ref{deltas}) we are left with a function
which depends only linearly on $m_1$ and $m_2$:
\begin{eqnarray*}
Z_N & = & \sum_\sigma \sum_{\mu_1 \in R_{N_1}  \atop \mu_2 \in R_{N_2} }
          \delta_{m_1, \mu_1} \delta_{m_2, \mu_2}
    \exp\big\{\frac{N}{2}(J_{11}\alpha^2(2m_1 \mu_1 - \mu_1^2)   +
              2J_{12}\alpha(1-\alpha)m_1\mu_2 + {}
           \nonumber\\
    & & {}+ J_{22}(1-\alpha)^2(2m_2 \mu_2 - \mu_2^2)) +  2J_{12}\alpha(1-\alpha)m_2\mu_1 +
     2J_{12}\alpha(1-\alpha)\mu_1\mu_2+
            \nonumber\\
    & & {} +  h_1N_1m_1 + h_2N_2m_2 \big \},
\end{eqnarray*}

\no and bounding above the Kronecker deltas by 1 we get
\begin{eqnarray} \label{deltas2}
Z_N & \leqslant
        & \sum_\sigma \sum_{\mu_1 \in R_{N_1}  \atop \mu_2 \in R_{N_2} }
              \exp\big\{-\frac{N}{2}(J_{11}\alpha^2\mu_1^2   + 2J_{12}\alpha(1-\alpha)\mu_1\mu_2 +J_{22}(1-\alpha)^2\mu_2^2)+
          \nonumber\\
 & &  + \big( (J_{11}\alpha \mu_1   + J_{12}(1-\alpha)\mu_2 +  h_1 ) N_1 m_1 +
      ( J_{12}\alpha\mu_1 + J_{22}(1-\alpha)\mu_2    +  h_2 ) N_2 m_2  \big) \big\}.
\nonumber\\
\end{eqnarray}

Since both sums are taken over finitely many terms, it is possible to exchange the order of the
two summation symbols, in order to carry out the sum over the spin configurations, which now
factorizes, thanks to the linearity of the interaction with respect to the $m$s. This way we get:

$$
Z_N \leqslant \sum_{\mu_1 \in R_{N_1} \atop \mu_2 \in R_{N_2} } G(\mu_1, \mu_2).
$$
where
\begin{eqnarray}\label{g}
G(\mu_1,\mu_2) &=&  \exp\big\{-\frac{N}{2}(J_{11}\alpha^2\mu_1^2   + 2J_{12}\alpha(1-\alpha)\mu_1\mu_2 +
	      J_{22}(1-\alpha)^2\mu_2^2) \big\}\cdot
          \nonumber\\
 & &   \cdot 2^{N_1}\big( \cosh \big( J_{11}\alpha \mu_1   + J_{12}(1-\alpha)\mu_2 +  h_1 \big) \big) ^{N_1}
          2^{N_2}\big( \cosh \big( J_{12}\alpha\mu_1 + J_{22}(1-\alpha)\mu_2    +  h_2 \big) \big) ^{N_2} \nonumber\\
\end{eqnarray}



Since the summation is taken over the
ranges   $R_{N_1}$  and $R_{N_2}$, of cardinality $N_1+1$ and $N_2+1$, we get that the total number of terms
is $(N_1+1)(N_2+1)$. Therefore
\begin{eqnarray}
Z_N & \leqslant   & (N_1+1)(N_2+1) \sup_{\mu_1, \mu_2} G,
 \end{eqnarray}

\no which leads to the following upper bound for $P_N$:
\begin{eqnarray} \label{deltas3}
P_N & = & \frac{1}{N} \ln Z_N  \leqslant \frac{1}{N} \ln (N_1+1) + \frac{1}{N} \ln (N_2+1) +
\frac{1}{N} \ln \sup_{\mu_1, \mu_2} G \; .
\end{eqnarray}


Now defining the $N$ independent function
\begin{eqnarray}
p_{UP}(\mu_1,\mu_2) =\frac{1}{N} \ln G  & =  & \ln 2 -\frac{1}{2}(J_{11}\alpha^2\mu_1^2   + 2J_{12}\alpha(1-\alpha)\mu_1\mu_2 +J_{22}(1-\alpha)^2\mu_2^2)  +
          \nonumber\\
 & & +   \alpha \ln \cosh \big( J_{11}\alpha \mu_1   + J_{12}(1-\alpha)\mu_2 +  h_1 \big) \big)+
         \nonumber\\
 & &        (1-\alpha) \ln \cosh  \big( J_{12}\alpha\mu_1 + J_{22}(1-\alpha)\mu_2    +  h_2 \big) \big),
\end{eqnarray}

\no the thermodynamic limit gives:
\begin{equation}
\limsup_{N \rightarrow \infty} P_N \leqslant  \sup_{\mu_1, \ \mu_2} p_{UP}(\mu_1,\mu_2).
\end{equation}



We can summarize the previous computation into the following:

\begin{lemma}
Given a Hamiltonian as defined in (\ref{Ham}), and defining the pressure per particle as $p_N=\frac 1 N \ln Z$,
given parameters $J_{11}, J_{12}, J_{22}, h_1, h_2$ and $\alpha$,
the following inequality holds:
$$
				\limsup_{N\rightarrow \infty} p_N \leqslant \sup_{\mu_1, \mu_2} p_{UP}
$$
\no where
\begin{eqnarray}
p_{UP} & =  & \ln 2 -\frac{1}{2}(J_{11}\alpha^2\mu_1^2   + 2J_{12}\alpha(1-\alpha)\mu_1\mu_2 +J_{22}(1-\alpha)^2\mu_2^2)  +
          \nonumber\\
 & & +   \alpha \ln \cosh \big( J_{11}\alpha \mu_1   + J_{12}(1-\alpha)\mu_2 +  h_1 \big) \big)+
         \nonumber\\
 & &        (1-\alpha) \ln \cosh  \big( J_{12}\alpha\mu_1 + J_{22}(1-\alpha)\mu_2    +  h_2 \big) \big),
\end{eqnarray}
\no and $(\mu_1, \mu_2) \in [-1,1]^2$.
\end{lemma}

\subsection{Lower bound}

The lower bound is provided by exploiting the well-known Gibbs entropic
variational principle (see \cite{ruelle}, pag. 188). In our case, instead of
considering the whole space of {\it ansatz} probability distributions considered in
\cite{ruelle}, we shall restrict to a much smaller one, and use the upper bound derived in the
last section in order to show that the lower bound corresponding to the restricted space
is sharp in the thermodynamic limit.

The mean-field nature of our Hamiltonian allows us to restrict the variational problem
to a two-degrees of freedom product measures represented through the non-interacting
Hamiltonian:
$$
\tilde H=-r_1\sum_{i=1}^{N_1}\xi_i-r_2\sum_{i=1}^{N_2}\eta_i,
$$
%
and so, given a Hamiltonian $\tilde H$, we define the ansatz Gibbs state corresponding to it as
$f(\sigma)$ as:
 \begin{equation*}
    \tilde\omega(f)=\frac{\sum_{\sigma}f(\sigma) e^{-\tilde H(\sigma)}}{\sum_{\sigma}e^{-\tilde H(\sigma)}}
    \end{equation*}

In order to facilitate our task, we shall express the variational principle of \cite{ruelle} in the
following simple form:

\begin{proposition}\label{varprin}
Let a Hamiltonian $H$, and its associated partition function $\displaystyle{Z=\sum_{\sigma}e^{-H}}$
be given. Consider an arbitrary trial Hamiltonian $\tilde H$ and its associated partition function $\tilde Z$.
The following inequality holds:
\begin{equation}\label{var ineq}
\ln Z \geqslant \ln \tilde Z - \tilde \omega (H) + \tilde {\omega} (\tilde H) \; .
\end{equation}
Given a Hamiltonian as defined in (\ref{Ham}) and its associated pressure per particle
$p_N=\frac 1 N \ln Z$,
the following inequality follows from {\rm (\ref{var ineq})}:
\begin{equation}\label{plow}
\liminf_{N\rightarrow \infty} p_N \geqslant \sup_{\mu_1, \mu_2} p_{LOW}
\end{equation}
\no where
\begin{eqnarray}
p_{LOW}(\mu_1, \mu_2)  &  = & \frac{1}{2}(J_{11}\alpha^2\mu_1 ^2+J_{22}(1-\alpha)^2\mu_2^2  +
            2J_{12}\alpha(1-\alpha) \mu_1 \mu_2) +
            \nonumber\\
    & & +    \alpha h_1 \mu_1 + (1-\alpha)h_2\mu_2+
            \nonumber\\
    & & +\alpha (- \frac{1+\mu_1}{2}\ln (\frac{1+ \mu_1}{2})  - \frac{1-\mu_1}{2}\ln (\frac{1- \mu_1}{2}))+
           \nonumber\\
  & &  + (1-\alpha)( - \frac{1+\mu_2}{2}\ln (\frac{1+ \mu_2}{2}) - \frac{1-\mu_2}{2}\ln (\frac{1- \mu_2}{2})).
\end{eqnarray}
\no and $(\mu_1, \mu_2) \in [-1,1]^2$.
\end{proposition}

\begin{proof}
The (\ref{var ineq}) follows straightforwardly from Jensen's inequality:
\begin{equation}
e^{\tilde\omega(-H+\tilde H)} \le \tilde \omega (e^{-H+\tilde H}) \; .
\end{equation}


It is convenient to express the Hamiltonian using the simbol $\xi$ for the spins of type $A$ and $\eta$ for
those of type $B$ as:
\begin{equation}\label{xieta_ham}
H(\sigma)=-\frac{1}{2N}(J_{11}\sum_{i,j}\xi_i\xi_j+2J_{12}\sum_{i,j}\xi_i\eta_j+J_{22}\sum_{i,j}\eta_i\eta_j)-h_1\sum_{i}\xi_{i}-h_2\sum_{i}\eta_{i} \; ;
\end{equation}

\no indeed its expectation on the trial state is
\begin{equation}
    \tilde\omega(H)=-\frac{1}{2N}(J_{11}\sum_{i,j}\tilde\omega(\xi_i\xi_j)+2J_{12}\sum_{i,j}\tilde\omega(\xi_i\eta_j)+
    J_{22}\sum_{i,j}\tilde\omega(\eta_i\eta_j))-h_1\sum_{i}\tilde\omega(\xi_{i})-h_2\sum_{i}\tilde\omega(\eta_{i})
\end{equation}

\no and a standard computation for the moments leads to
\begin{eqnarray}
    \tilde\omega(H) & = & -\frac{N}{2}(J_{11}(\alpha^2-\alpha/N)(\tanh r_1) ^2+ J_{11}\alpha/N+J_{22}
    ((1-\alpha)^2-(1-\alpha)/N)(\tanh r_2)^2  +
    \nonumber\\
 & &   + J_{22}(1-\alpha)/N + 2J_{12}\alpha(1-\alpha) \tanh r_1 \tanh r_2) - N\alpha h_1\tanh r_1 -
    N(1-\alpha)h_2\tanh r_2.
    \nonumber\\
\end{eqnarray}

Analogously, the Gibbs state of $\tilde H$ is:
\begin{equation*}
    \tilde\omega(\tilde H)  =   -N \alpha r_1 \, \tanh r_1 - N(1-\alpha)r_2\tanh r_2,
\end{equation*}

\no and the non interacting partition function is:
$$
\tilde Z_N= \sum_{\sigma} e^{- \tilde H(\sigma)} =
2^{N_1}(\cosh r_1)^{N_1} + 2^{N_2}(\cosh r_2)^{N_2}
$$

\no which implies that the non-interacting pressure gives
$$
\tilde p_N = \frac 1 N \ln \tilde Z_N = \ln 2 + \alpha \ln \cosh r_1 +  (1-\alpha) \ln \cosh r_2
$$

So we can finally apply Proposition (\ref{var ineq}) in order to find a lower bound for the pressure
$p_N=\displaystyle{\frac 1 N} \ln Z_N$:
\begin{eqnarray}
    p_N=\frac{1}{N}\ln Z_N  \geqslant \frac{1}{N}\   \Big( \ln \tilde Z_N
    - \tilde \omega (H) + \tilde {\omega} (\tilde H) \Big)
\end{eqnarray}

\no which explicitly reads:
\begin{eqnarray}\label{nonprecise}
    p_N=\frac{1}{N}\ln Z_N  & \geqslant &  \ln 2 + \alpha \ln \cosh r_1 + (1-\alpha) \ln \cosh r_2 + {}
          \nonumber\\
    & &  {} + \frac{1}{2}(J_{11}\alpha^2(\tanh r_1) ^2+J_{22}(1-\alpha)^2(\tanh r_2)^2  +2J_{12}\alpha(1-\alpha)
    \tanh r_1 \tanh r_2) +
           \nonumber\\
    & & + \alpha h_1\tanh r_1 +(1-\alpha)h_2\tanh r_2  -\alpha r_1\tanh r_1 - (1-\alpha)r_2\tanh r_2
          \nonumber\\
    & &  + J_{11}\alpha/2N +J_{22}(1-\alpha)/2N-J_{11}\alpha (\tanh r_1) ^2/N-J_{22}(1-\alpha)(\tanh r_2)^2/N.
          \nonumber\\
\end{eqnarray}

Taking the lim inf over $N$ and the supremum in $r_1$ and $r_2$ of the left hand side we get the (\ref{plow})
after performing the change of variables $\mu_1=\tanh r_1$ and $\mu_2=\tanh r_2$.

\end{proof}

\subsection{Exact solution of the model}
	
Though the functions $p_{LOW}$ and $p_{UP}$ are different, it is easily checked that they share the same local suprema.
Indeed, if we differentiate both functions with respect to parameters $\mu_1$ and $\mu_2$, we see that the extremality conditions are given in both cases by the Mean Field Equations:
\begin{equation}\label{MFE}
\left\{ \begin{array}{lll}
\mu_1 & = & \tanh(J_{11}\alpha \mu_1   + J_{12}(1-\alpha)\mu_2 +  h_1 )
\\
\mu_2 & =  & \tanh(J_{12}\alpha\mu_1 + J_{22}(1-\alpha)\mu_2    +  h_2 )
\end{array} \right.
 \end{equation}

If we now use these equations to express $\tanh^{-1} \mu_i$ as a function of $\mu_i$ and we substitute back
into $p_{UP}$ and $p_{LOW}$ we get the same function:
\begin{equation}\label{presfunc}
p(\mu_1, \mu_2)  = -\frac{1}{2}(J_{11}\alpha^2\mu_1^2   + 2J_{12}\alpha(1-\alpha)\mu_1\mu_2 +J_{22}(1-\alpha)^2\mu_2^2)  +
    -  \frac{1}{2}\alpha\ln \displaystyle{\frac{1-\mu_1^2}{4}}- \frac{1}{2}(1-\alpha) \ln \displaystyle{\frac{1-\mu_2^2}{4}}.
\end{equation}

Since this function returns the value of the pressure when the couple ($\mu_1$, $\mu_2$) corresponds to an extremum, and this is the same both for $p_{LOW}$ and $p_{UP}$, we have proved the following:

\begin{theorem}
Given a hamiltonian as defined in (\ref{xieta_ham}), and defining the pressure per particle as
$\displaystyle{p_N=\frac 1 N \ln Z}$,
given parameters $J_{11}, J_{12}, J_{22}, h_1, h_2$ and $\alpha$, the thermodynamic limit
$$
\lim_{N\rightarrow \infty} p_N = p
$$
\no of the pressure exists,
and can be expressed in one of the following equivalent forms:
      \newcounter{Lcount}
	\begin{list}{\alph{Lcount})}
	{\usecounter{Lcount}
	\setlength{\rightmargin}{\leftmargin}}
		\item $\displaystyle{p = \sup_{\mu_1, \mu_2} \ p_{LOW}(\mu_1, \mu_2)}$
		\item $\displaystyle{p = \sup_{\mu_1, \mu_2} \ p_{UP}(\mu_1, \mu_2)}$
    \end{list}
\end{theorem}

\section{Preliminary analytic result}\label{maxima}

Though analysis cannot solve our problem exactly, it can tell us a what to expect 
when we solve it numerically.
In particular, in this section we shall prove that, for any choice of the parameters, the
total number of local maxima for the function $p(\mu_1, \mu_2)$ is less or equal to five.

We recall that the mean field equations for our two-population model are:
\begin{displaymath}
\left\{ \begin{array}{lll}
\mu_1 & = & \tanh(J_{11}\alpha \mu_1   + J_{12}(1-\alpha)\mu_2 +  h_1 )
\\
\mu_2 & =  & \tanh(J_{12}\alpha\mu_1 + J_{22}(1-\alpha)\mu_2    +  h_2 )
\end{array}, \right.
 \end{displaymath}

\no and correspond to the stationarity conditions of $p(\mu_1, \mu_2)$. So, a subset of
solutions to this system of equations are local maxima, and some among them correspond
to the thermodynamic equilibrium.


These equations give a two-dimensional generalization of the Curie-Weiss mean field equation.
Solutions of the classic Curie-Weiss model can be analysed by elementary geometry:
in our case, however, the geometry is that of 2 dimensional maps, and it pays to recall
that Henon's map, a simingly harmless 2 dimensional diffeomorhism of $\R^2$, is known to exhibit
full-fledged chaos. Therefore, the parametric dependence of solutions, and in particular the number
of solutions corresponding to local maxima of $p(\mu_1, \mu_2)$, is  in no way
apparent from the equations themselves.

We can, nevertheless, recover some geometric features from the analogy with one-dimensional picture.
For the classic Curie-Weiss equation, continuity and the Intermediate Value Theorem from elementary calculus
assure the existence of at least one solution. In higher dimensions we can resort to the analogous result, 
{\it Brouwer's Fixed Point Theorem}, which states that any continuous map on a topological closed ball
has at least one fixed point.
This theorem, applied to the smooth map $R$ on the square $[-1,1]^2$, given by
\begin{displaymath}
\left\{ \begin{array}{lll}
R_1(\mu_1,\ \mu_2) & = & \tanh(J_{11}\alpha \mu_1   + J_{12}(1-\alpha)\mu_2 +  h_1 )
\\
R_2(\mu_1,\ \mu_2) & =  & \tanh(J_{12}\alpha\mu_1 + J_{22}(1-\alpha)\mu_2    +  h_2 )
\end{array} \right.
 \end{displaymath}
establishes the existence of at least one point of thermodynamic equilibrium. 

We can gain further information by considering the precise form of the equations:
by inverting the
hyperbolic tangent in the first equation,  we can $\mu_2$ as a function of $\mu_1$, and vice-versa for
the second equation.
Therefore, when $J_{12}\neq0$ we can rewrite the equations in the following fashion:
\begin{eqnarray} \label{gamma1}
\left\{ \begin{array}{lll}
\mu_2 & = & \displaystyle{ \frac{1}{J_{12}(1-\alpha)} (\tanh^{-1} \mu_1 -  J_{11}\alpha \mu_1- h_1 ) }
\\
\mu_1 & = & \displaystyle{\frac{1}{J_{12}\alpha} (\tanh^{-1}\mu_2 -  J_{22}(1-\alpha) \mu_2- h_2 ) }
\end{array} \right.
 \end{eqnarray}

Consider, for example, the first equation: this defines a function $\mu_2(\mu_1)$, and we shall
call its graph {\it curve $\gamma_1$}. Let's
consider the second derivative of this function:
$$
\frac{\partial^2\mu_2}{\partial \mu_1 ^2}=-\frac{1}{J_{12}(1-\alpha)} \cdot \frac{2 \mu_1}{(1-\mu_1^2)^2}.
$$

\begin{figure}
    \centering
    \includegraphics[width=8 cm]{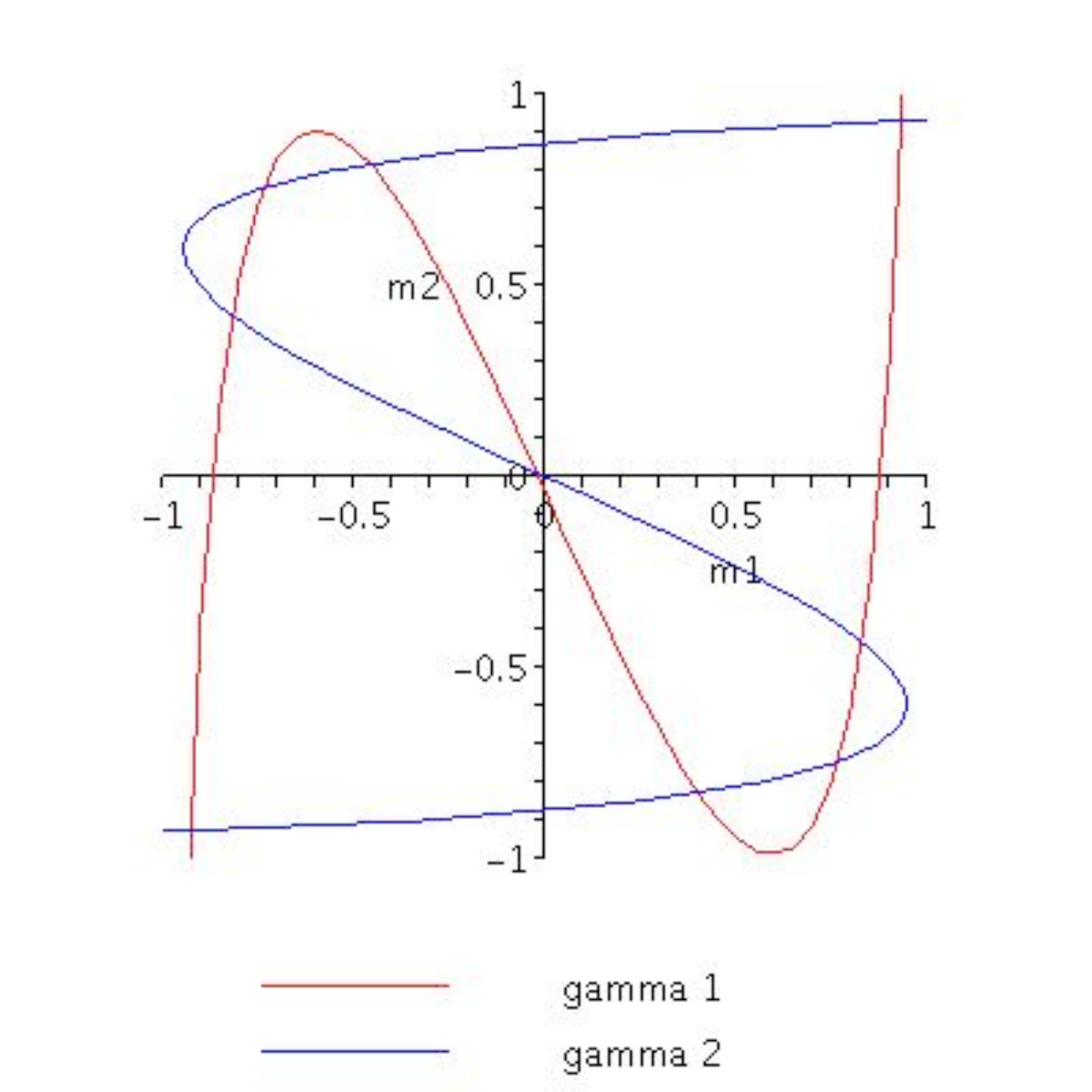}
    \caption{The crossing points correspond to solutions of the mean field equations}
    \label{9points}
\end{figure}

We see immediately that this second derivative is strictly increasing, 
and that it changes sign exactly at zero.
This implies that $\g_1$ can be divided into three monotonic pieces, each having
strictly positive {\it third} derivative as a function of $\mu_1$.
The same thing holds for the second equation, which defines a function $\mu_1(\mu_2)$,
and a corresponding {\it curve $\gamma_2$}.
An analytical argument easily establishes that there exist at most $9$
crossing points of $\g_1$ and $\g_2$  (for convenience we shall label the three monotonic pieces
of $\g_1$ as $I$, $II$ and $III$, from left to right): since $\g_2$, too, has a strictly positive third derivative,
it follows that it intersects each of the three monotonic pieces of $\g_1$ at most three times, and this
leaves
the number of intersections between $\g_1$ and $\g_2$ bounded above by 9 (see an example
of this in Figure \ref{9points}).

By definition of the mean field equations, the stationary points of the pressure correspond to
crossing points of $\gamma_1$ and $\gamma_2$. 
Furthermore, common sense tells us that not all of these stationary points can be local maxima.
This is indeed true, and it is proved by the following:

\begin{proposition}\label{5max}
The function $p(\mu_1, \mu_2)$ admits at most 5 maxima.
\end{proposition}

To prove \ref{5max} we shall need the following:

\begin{lemma}\label{adjpoints}
Say $P_1$ and $P_2$ are two crossing points linked by a monotonic piece of one of the two functions
considered above. Then at most one of them is a local maximum of the pressure $p(\mu_1, \mu_2)$.
\end{lemma}

\begin{proofof}{Lemma \ref{adjpoints}}
The proof consists of a simple observation about the meaning of our curves.
The mean field equations as stationarity conditions for the pressure,
so each of $\gamma_1$ and $\gamma_2$
are made of points where one of the two components of the gradient
of $p(\mu_1, \mu_2)$ vanishes. Without loss of generality assume that $P_1$ is a maximum,
and that the component that vanishes on the piece of curve that links $P_1$ to $P_2$ is
$\ds{\frac{\partial p}{\partial \mu_1}}$.

Since $P_1$ is a local maximum, $p(\mu_1, \mu_2)$ locally increases on the piece of curve $\gamma$.
On the other hand, the directional derivative of $p(\mu_1, \mu_2)$ along $\gamma$  is given by
$$
\mathbf{\hat t} \cdot \nabla p
$$

\no where $\mathbf{\hat t}$ is the unit tangent to $\gamma$. Now we just need to notice that
by assumptions for any
point in $\gamma$ $\mathbf{\hat t}$ lies in the same quadrant, while $\nabla p$ is vertical with
a definite verse.
This implies that the scalar product giving directional derivative is strictly non-negative over all $\gamma$,
which prevents $P_2$ form being a maximum.
\end{proofof}

\begin{proofof}{Proposition \ref{5max}}
The proof considers two separate cases:

\begin{itemize}
\item[a)] All crossing points can be joined in a chain by using monotonic pieces of curve such as the one
          defined in the lemma;
\item[b)] At least one crossing point is linked to the others only by non-monotonic pieces of curve.
\end{itemize}

In case a), all stationary can be joined in chain in which no two local maxima can be nearest neighbours,
by the lemma.
Since there are at most 9 stationary points, there can be at most 5 local maxima.

For case b) assume that there is a point, call it $P$, which is not linked to any
other point by a monotonic piece of curve.
Without loss of generality, say that $P$ lies on $I$ (which, we recall, is defined as the leftmost
monotonic piece of $\gamma_1$). By assumption, $I$ cannot contain other crossing points
apart from $P$, for otherwise $P$ would be monotonically linked to at least one of them, contradicting 
the assumption.
On the other hand,  each of $II$ and $III$  contain at most $3$ stationary points, and, by Lemma \ref{adjpoints}, at most $2$ of these are maxima. So we have at most $2$ maxima on each of
$II$ and $III$, and and at most 1 maximum on $I$,  which leaves the total bounded above by $5$.
The cases in which $P$ lies on $II$, or on $III$, are proved analogously, giving the result.

\end{proofof}

%
%

\section{Comments}

The considered model generalises models which arise naturally as
approximations of various problems in theoretical physics.
Furthermore, the upcoming study of
social phenomena by statistical mechanics methods
provides another importance source of interest for a model describing long-range
interactions between two homogeneous populations.

In \cite{cgm} we show that it is possible to
give a cultural-contact interpretation to the model presented here, and thanks to the mathematical 
results just derived, to provide non-trivial information about its regimes.

 It is not known at present which is the exact mathematical structure underlying social
 networks. However, it is well-accepted that interactions must be of the ``small world''
 type predicted in \cite{milgram}, at least to some degree. We plan to return
 on those topics in future works.

{\bf Acknowledgments}. We thank Cristian Giardin\`a and Christof K\"ulske for many interesting 
discussions.

%

%
%
%
%
%
%
%
%
%

\end{document}